\documentclass[twocolumn]{ceurart}

\begin{document}
\copyrightyear{2021}
\copyrightclause{Copyright for this paper by its authors.
  Use permitted under Creative Commons License Attribution 4.0
  International (CC BY 4.0).}

\conference{ Causality in Search and Recommendation (CSR) and Simulation of Information Retrieval Evaluation (Sim4IR) workshops at SIGIR, 2021 
Editors of the proceedings (editors): Krisztian Balog, Xianjie Chen, Xu Chen, David Maxwell, Paul Thomas, Shuo Zhang, Yi Zhang, and Yongfeng Zhang}

\title{State of the Art of User Simulation approaches for conversational information retrieval}

\author[1]{Pierre Erbacher}[%
email=pierre.erbacher@lip6.fr
]

\author[1]{Laure Soulier}[%
email=laure.soulier@lip6.fr,
]

\address[1]{Sorbonne Université, CNRS, LIP6,
 F-75005 Paris - France}


\author[2]{Ludovic Denoyer}[%
email=denoyer@fb.com,
]
\address[2]{Facebook AI research}

\begin{abstract}
  Conversational Information Retrieval (CIR) is an emerging field of Information Retrieval (IR) at the intersection of interactive IR and dialogue systems for open domain information needs. In order to optimize these interactions and enhance the user experience, it is necessary to improve IR models by taking  into account sequential heterogeneous user-system interactions. Reinforcement learning has emerged as a paradigm particularly suited to optimize sequential decision making in many domains and has recently appeared in IR. However, training these systems by reinforcement learning  on users is not feasible. One solution is to train IR systems  on user simulations that model the behavior of real users.  Our contribution is twofold: 1) reviewing the literature on user modeling and user simulation for information access, and 2) discussing the different research perspectives for user simulations in the context of CIR.
\end{abstract}


\maketitle{}

\section{Introduction}

Conversational information retrieval (CIR) is a well-established research field in the Information Retrieval (IR) community  \cite{10.1145/3274784.3274788}.
It is at the intersection of many application domains (interactive IR, conversational systems, question answering) but differs in many aspects.
1) Unlike interactive IR \cite{10.1145/3397271.3401424} that is also interested in user feedback, CIR's objective is to find the relevant information in a more natural way (i.e., interactions in natural language). 
2) It is also different from task-oriented conversation systems \cite{DBLP:conf/iclr/BordesBW17} which are guided in a particular application area (i.e., travel booking). A CIR system is a specific application case of task-oriented conversation systems, but is different because the search is performed in an open domain.
3) Finally, very close to question answering systems \cite{bordes-etal-2014-question} in which questions are expressed in natural language, CIR focuses on exploratory or complex information needs, on open fields, not specifically centered on a specific entity, question or document as in Q\&A systems.

The objective of a CIR system is to help the user in his/her interactions with a search engine through a conversational tool. This whole search engine/conversational system will aim to: 1) respond to a complex information need  expressed in natural language and/or through various feedbacks (e.g., sequences of reformulation of requests, clicks on documents, conversations in natural language)  and 2) anticipate and guide the user in his/her sequence of search actions using, e.g.,  query clarification or document ranking \cite{Azz2018,10.1145/3020165.3020183}. This requires CIR systems to interact with users to better understand and/or refine their information needs in a long term objective. 


Most recent models focus on the interactive IR aspect that builds a user profile or session modeling \cite{Sordoni2015ASuggestion,10.1145/3397271.3401181,10.1145/3331184.3331221}. However, they do not support on a long-term, sequential and incremental decision strategy. The latter has been outlined as essential  for conversational recommendation systems  \cite{Sun2018ConversationalSystem}. A growing interest in the community is to design IR systems that embed strategies to best optimize these search sequences \cite{DBLP:journals/corr/abs-1909-12425,10.1145/2600428.2602297}, in particular through interactions in natural language with users  \cite{Zhang2020DeepAdvances}. 
Reinforcement learning has been shown to be particularly effective in optimizing sequential decision making  \cite{sutton1998rli} because it can learns a policy that maximizes a long term cumulative reward using successive interactions with its environment. The idea of using reinforcement learning to optimize an IR policy has become a hot topic \cite{Zhang2020DeepAdvances,10.1145/3397271.3401099}. 
For IR, policies would be learned to optimize in the long term the sequences of user interactions with the CIR system according to different criteria : reducing/increasing the number of interactions with the user, increasing the diversity of retrieved information, predicting future needs, guiding users actions, etc.
Ideally, CIR system policies should train with real users. In practice, this is impossible because reinforcement learning algorithms are currently very inefficient: it would take millions of interactions with users to see the emergence of a correct policy. 

Training directly on data limits the exploration to scenarios collected from real users, so CIR systems can only visit a very small part of the state space. In addition, even if a lot of search trajectories (sequence of interactions between user and system) are collected, only a few would be optimal \cite{Kreyssig}.

One of the solutions is to train CIR systems with user simulations (US). US generate synthetic data close to data generated by real users. This makes it possible to : 1) estimate the most probable actions of users, particularly in new scenarios and 2) best optimize CIR systems policies by exploring trajectories not contained in collected scenarios. In this article we are interested in the problem of user modeling and simulation in different contexts of access to information (e.g., IR, recommendation, conversational systems).

\begin{enumerate}
    \item We review the different approaches for modeling and simulating user for information access and goal oriented conversational systems (Section 2). Modeling and simulation are often confused in the literature. For a source system, in our case the user whose behavior is unknown, the associated model is a set of instructions, rules, markov chain or equations constructed from the observed data coming from this user \cite{Zeigler1984TheoryBooks}. Simulation is an agent that generates behaviors by following the rules of a model. Together, the goal is to emulate user behavior \cite{Zeigler1984TheoryBooks}.

    \item We discuss state-of-the-art SU evaluation, the difficulty being to measure whether the simulated behavior is realistic (Section ~ 3).  
    
    \item Finally, we discuss the research perspectives to build user simulations for CIR (Section 4). 
\end{enumerate}


\section{Model and simulate user behavior}

 

The IR community produced lot of work in studying and modelling users behaviors either based on cognitive or statistical approaches. First user modeling approaches in IR were based on cognitive approaches. For instance, Belkin \cite{Belkin1984CognitiveTransfer} produces a description of the users characterized by an objective or a problem as well as a knowledge of the world. He also describes the conditions necessary for an IR system to be effective with ambiguous queries. Many derivatives were then proposed  (e.g., ISP \cite{Kuhlthau:1991},  IS\&R \cite{Ingwersen:2005b},  Ellis \cite{Ellis:1989}). The second line of work  mainly remains on statistical models describing users' click behavior and satisfaction during search sessions  \cite{Craswell2008AnModels,Chapelle2009ARanking,Dupret2008AObservations,Chuklin2013ModelingPage}. Beyond being used for building predictive models (from the IR system point of view), these user models serve also as basis of user simulations. In what follows, we will focus on the literature review of user simulation models. Please note that since CIR is a young research domain, most of the cited work below belongs to orthogonal and related fields, namely recommendations and dialog. 

Historically, US have been extensively studied for conversational systems \cite{Eckert1997UserEvaluation, Komatani2005UserGuidance, Pietquin2004A1, Schatzmann2006AStrategies, Scheffler2000ProbabilisticDialogues}. One of the first attempts is to use statistical user models as a simulator \cite{Eckert1997UserEvaluation}. This simulates the users from the appearance probabilities of a sentence according to the previous sentence. This simulation is limited, because it only takes into account the previous response of the system, the probabilities suffer from the lack of data and moreover it is not guided by a dialogue objective. To add consistency to users' responses, Scheffler et al. \cite{Scheffler2000ProbabilisticDialogues} introduces into its simulator a dialogue objective, drawn from a list of possible objectives. The model follows an imposed dialogue objective: the probabilities are conditioned by the objectives of the real users in the corpus.

Komatani et al.\cite{Komatani2005UserGuidance} bring diversity by conditioning the probability of utterance on the user's ability to use a search system as well as its degree of urgency. From annotated data, a decision tree is used to classify the sentences of the corpus according to the urgency and the competence of the users. For the simulation, a degree of urgency is drawn and imposed on the model for the generation of sentences.
One way to implement rule based US is to use agenda, as in \cite{Schatzmann2007Agenda-BasedSystem}. In this work, the authors build a conversational US using constraints and rules which make it possible to explicitly encode the user's state as well as his objective. In addition to generating sentences according to a certain user model, the simulator is assigned a dialogue objective as well as a dialogue strategy (Agenda) which conditions the sentences generated by the model. 
  
Agenda-based approaches are widely used to train conversational recommendation systems because they allow to build dialogue strategies for close domain using simple rules \cite{Hou2019ADialogue,Li2016A}. Zhang et al. \cite{Zhang2020EvaluatingSimulation} show that US Agendas can be particularly realistic: when recommending movies, 36\% of human reviewers mistook an US for a real user and 23\% of reviewers were undecided.

Li et al. \cite{Li2016A} propose a more realistic hybrid model that relies on the agenda to maintain user dialogue strategy but which takes advantage of the power of language models \textit{seq2seq} to improve the natural language generation and language understanding. Asri et al. \cite{Asri2016ASystems} and Kreyssig et al. \cite{Kreyssig} propose a neural model that breaks free from the agenda by using \textit{seq2seq} language models to generate and interpret natural language. On the other hand, they keep the generation of objectives (constraints / requests) at the semantic level. To simulate the dynamic behavior of the user, the constraints are drawn independently according to a probability observed in the dataset.

Some work relies on deep learning to simulate users from data \cite{Chandramohan2011UserLearning,chen2019generative,Zhao2019SimulatingRecommendations}. For example, Zhao et al. \cite{Zhao2019SimulatingRecommendations} learn a user simulation which provides information on the relevance of the recommendations made by the system. They use an adversarial generative approach to converge to a US. The discriminator must differentiate between simulated and real feedback based on user search history. The generator learns to create feedbacks based on a history of interactions.
By relying on reinforcement learning approaches, Chen et al. \cite{chen2019generative} propose to jointly model the user click policy and the reward function using a generative approach. Indeed, the user policy is optimized to generate the actions which maximize the reward signal. In parallel, the reward function has the role of the discriminator, it is optimized to minimize the difference in reward between the action taken by the user policy and the action taken by a real user observed in the database.
A data driven approach is based on reverse reinforcement learning, the objective of which is to learn the reward function from the behavioral observations of another (often human) agent \cite{Russell1998LearningAbstract}. This approach contrasts with previous models which relied on explicitly defined reward functions. In this context, Chandramohan et al.  \cite{Chandramohan2011UserLearning} propose to use inverse reinforcement learning to find the reward function describing user policies observed in data. An experiment is carried out using conversational data in the field of tourist information.

Some US relies on active learning. Because learning a user model requiring a large amount of data. For this, Zhang et al. \cite{Zhang2019BudgetedSystems} proposes to include sparsely human feedback in the learning loop. In this model, the user model is learned jointly with the recommendation system. The goal is to optimize training to limit human intervention. For this, the user model is allocated a limited budget. At each turn, the model can choose how to allocate her credits to respond to the recommendation system: 0 credit to respond using the learned policy, 1 credit to access a human response, 2 credit to collect a human discussion - human.
The user model learns from the new data collected. At the start of each conversation, a goal generator pulls in unsuccessful or unexplored goals. Despite a parsimonious use of human interventions, including humans in the learning loop remains time-consuming.

\section{Evaluation of User Simulation}

As explained earlier, one of the objectives of US is to evaluate \cite{Eckert1997UserEvaluation} information retrieval systems. It is therefore needed to ensure the quality of the US used for the evaluation. 
According to \cite{Pietquin2008ASimulations}, a good metric, to evaluate US, should be able to measure US quality independently of the task and should be correlated with real user evaluations.
Following \cite{Pietquin2008ASimulations}, we distinguish two types of evaluation. 


Direct evaluation directly measures US through metrics. For example, an evaluation approach consists in measuring whether the policy has achieved its objective, the quality of sentences generated in a dialogue system thanks to BLEU / RED / METEOR measurements \cite{papineni-etal-2002-bleu,Lin2004ROUGE:Summaries,Banerjee2005METEOR:Judgments,Sordoni2015AResponses}, likely click behavior measured by perplexity \cite{Dupret2008AObservations}). However, in the context of dialogue systems, \cite{liu-etal-2016-evaluate} demonstrates that automatic metrics that assess word overlap with reference sentences like BLUE, RED and METEOR have very little correlation with judgment human, in particular because there is a great diversity of valid answers for each round of conversation \cite{liu-etal-2016-evaluate,Zhang2020EvaluatingSimulation}.
 Human evaluation is also used to evaluate the quality of the responses of a user simulation. For example, Zhang et al. \cite{Zhang2020EvaluatingSimulation} observe the number of times human evaluators cannot differentiate between real user responses and US-generated dialogues to compare the quality between US.

Indirect evaluation measures the impact of a US on the performance of the retrieval system \cite{Schatzmann2006AStrategies}.
Shi et al. \cite{shi-etal-2019-build} assess the realism of its user model thanks to direct and indirect human annotation. For direct evaluation, they directly judge the quality of sentences from different US through different criteria: the quality of the language (such as grammar), the consistency of the sentences between the different turns as well as the diversity of the sentences. The evaluators also give an overall rating of the simulations. The indirect evaluation of the US is measured through the performance of the recommender systems that have interacted with the US. Human reviewers analyze the quality of the recommendation. They rate their satisfaction, the naturalness of the conversations and the effectiveness of the recommendations.

This section summarizes how human annotations and automatic metrics are used in the literature to assess US quality via direct and indirect evaluations. \cite{Zhang2020EvaluatingSimulation} shows that assessing IR systems using US is possible. There are still lot of work to be done to have reliable and realist user simulation in order to benchmark IR systems.

\section{Research perspectives}



In this section, we discuss the research perspectives on user simulation models for CIR.

\textbf{Towards multi-modal user simulation models for open domains (CIR).} 
 To generate more realistic search sessions, US should be able to express complex queries on many different domains and simulate document selection (click). Currently, conversational US do not embed user click models. 
 However, clicks are present in many IR scenarios for example when the system is returning keywords for disambiguating the query or selecting document. Generating complete user behaviors in the form of sequences of clicks and requests or dialogue seems essential to have realistic evaluation frameworks for CIR systems. To overcome the lack of data, one idea would be to incorporate existing click models \cite{Dupret2008AObservations} into US \cite{Schatzmann2007Agenda-BasedSystem}. Another idea is to jointly learn the language model and the user click model. Simulating both conversations and clicks seems essential for learning CIR systems that return a list of documents. 
Also, CIR imposes to be able to evolve on a variety of domains (open domains). The vast majority of US are built for conversational recommendation systems (CRS) for a specific domain, for example,(e.g., restaurant suggestion). 
CIR systems need to offer services covering a growing number of areas.
A first step towards this open domain would be to benefit from the research in multi-domain CRS \cite{Rastogi2019TowardsDataset}, with the publication of several multi-domain datasets \cite {Rastogi2019TowardsDataset, Budzianowski2018MultiWOZModelling}.\\

\textbf{Towards diversity-driven user simulation}

One of the difficulties brought by neural models is to generate diversity in behavior. Indeed, such models learn to satisfy the average behaviors observed in the data. However, there is a strong variability in user behavior \cite{White2007InvestigatingSearch}. Some are more persistent or curious in their research, while others give up quickly. In addition, the context and world knowledge is different between users. Learning a single user model on data does not reflect this diversity of users. Recently proposed US, model one behavior from data generated by many real users. One of the ideas would be to generate realistic user policies but with different styles. One method is to regroup and model by hand the different users according to a certain number of criteria (curiosity, perseverance, world knowledge, expertise, ...) such as \cite{Komatani2005UserGuidance}. A second method is to extract automatically these information from a corpus of conversations. Some works propose to use US with attributes and characteristics (persona) \cite{Li2016AModel,Balog2019PersonalAgenda} to obtain both diversity and consistency in requests and dialogues. Li et al. \cite{Li2016AModel} used user embedding encoding speaker specific information (persona) such as age, dialect, context, gender and other personal information that influences generated responses. Balog et al. \cite{Balog2019PersonalAgenda} proposed a more convenient way to encode information about users persona using knowledge graph. Ideally, CIR should use knowledge they have about the user in order to optimize this user's search session. In a cold start scenario, one possibility for the CIR is to probe users to extract a maximum number of information with a limited probing budget. We believe that using diversity driven US is essential to learn such probing policies.  \\ 



\textbf{Towards the establishment of benchmarks for CIR systems.} Currently, human annotators are the most reliable benchmark to assess CIR systems performances. But, human evaluation is  time-consuming to test the robustness of models for various information needs and behaviors. One of the solutions is to establish US to measure the performance of the various proposed CIR models. Note that \cite{Zhang2020EvaluatingSimulation} proposed this evaluation framework for CR systems. In addition, CIR systems should also be evaluated to on their generalization capacity:
\begin{itemize}
    \item To open domain conversation
    \item To various level of query complexity
    \item To various user behavior
\end{itemize}
For this, we believe that the benchmark should be a population of unseen US having different persona and behavior.

Learning users behaviors in CIR context directly from collected data might be limited because:
1) Collecting conversational search session data is computationally expensive. 2) Collected data is entangled with used the IR system. One idea is to build US with different level of abstractions. From the intention level to the natural language generation.
In order to facilitate future works, the IR community should have a common platform for building and testing US for sequential IR systems. In this direction, some environments simulating users have already been developed for learning recommendation system using sequential learning such as \cite{ie2019recsim,rohde2018recogym}.

\section{Conclusion}

This article focuses on CIR and above all on setting up a training and evaluation framework for user-centered approaches. Although human assessment is the most suitable, it is however too time-consuming for models based on deep learning or reinforcement learning. The objective of this paper is to synthesize the state of the art of user modeling and simulation approaches in various information access tasks to open up prospects for the implementation of CIR systems.


\section{Acknowledgments}
We would like to thank the ANR JCJC project SESAMS (Projet-ANR-18-CE23-0001) for supporting Pierre Erbacher and Laure Soulier from Sorbonne Univeristé in this work.

\bibliography{camera-ready}

\end{document}